\documentclass[letter,twocolumn]{jpsj2}

\newcommand{\epsfig}[1]{%
  \centerline{\resizebox{0.442\textwidth}{!}{\includegraphics{#1}}}%
}

\title{Randomness-Driven Quantum Phase Transition in
  Bond-Alternating Haldane Chain}

\author{Takayuki \textsc{Arakawa}\thanks{Present address: Speech
    Interface Technology Group, NEC Corporation, Kawasaki 211-8666,
    Japan.}, Synge \textsc{Todo}$^{1}$\thanks{E-mail address:
    wistaria@ap.t.u-tokyo.ac.jp} and Hajime \textsc{Takayama}}

\inst{Institute for Solid State Physics, University of Tokyo, Kashiwa
  277-8581, Japan\\ $^{1}$Department of Applied Physics, University of
  Tokyo, Tokyo 113-8656, Japan}

\abst{The effect of bond randomness on the spin-gapped ground state of
  the spin-1 bond-alternating antiferromagnetic Heisenberg chain is
  discussed.  By using the loop cluster quantum Monte Carlo method, we
  investigate the stability of topological order in terms of the
  recently proposed twist order parameter [M. Nakamura and S. Todo:
    Phys. Rev. Lett. {\bf 89} (2002) 077204].  It is observed that the
  dimer phases as well as the Haldane phase of the spin-1 Heisenberg
  chain are robust against a weak randomness, though the valence-bond-solid-like
  topological order in the latter phase is destroyed by introducing a
  disorder stronger than the critical value.}

\kword{Haldane chain, bond randomness, quantum phase transitions,
  random-singlet phase, quantum Monte Carlo, loop algorithm,
  topological order, twist order parameter}

\recdate{October 29, 2004}

\begin{document}
\maketitle

Disorder effects on low-dimensional quantum magnets have been
investigated extensively in recent theoretical studies.  In particular,
impurity effects on spin-gapped Heisenberg
antiferromagnets~\cite{MatsumotoYTT2002} have aroused much interest in
relation to the impurity-induced antiferromagnetic (AF) long-range order
observed experimentally in real materials~\cite{HaseKMSUS1995}.  It
has been established by recent numerical
simulations~\cite{YasudaTMT2001, YasudaTMT2002} that in two dimensions
or higher, there are two classes of disorder, that affect spin-gapped
states in essentially different ways.  Site dilution and bond
dilution are representatives of each class.  The former induces
localized moments around impurity sites.  There exist strong
correlations between such effective spins retaining the staggeredness
with respect to the original lattice, and therefore the
AF long-range order emerges by an infinitesimal
concentration of dilution.  In the bond-dilution case, on the other
hand, localized moments are always induced in pairs and they form a
singlet again by AF interactions through the two- or
three-dimensional shortest paths as long as the concentration of bond
dilution is smaller than a finite critical value.

In one dimension, since quantum fluctuations are much
stronger than those in higher-dimensional systems, novel quantum critical
phenomena are observed under disorder at the magnitude of coupling
constants (bond randomness).  Theoretically, the decimation
renormalization group (DRG) approaches have achieved great success in
predicting rich physics, such as the random-singlet (RS) phase for
spin-$\frac{1}{2}$ chains~\cite{DasguptaM1980, Fisher1994, HymanYBG1996}.  Recently,
this technique has been extended to higher-spin
cases~\cite{HymanY1997, MonthusGJ1997,
  Damle2002, SaguiaBC2002}, where two of the
main debates are on the robustness of the Haldane
gap~\cite{Haldane1983} against disorder and on the
presence of the spin-1 RS phase.  A number of numerical studies have
also been carried out~\cite{Nishiyama1998, Hida1999, TodoKT2000,
  CarlonLRI2004} to establish a quantitative phase diagram.
However, this problem has not been sufficiently clarified yet.  One of the
main difficulties in simulating random quantum systems is
the extremely wide energy scale that has to be taken into account.  Another
difficulty is the lack of appropriate physical quantity for effectively
discussing randomness-driven critical behavior.

In this Letter, we report the results of our quantum Monte Carlo (QMC)
simulation on the bond-alternating Haldane chain with bond randomness.
By using the recently proposed twist order
parameter~\cite{NakamuraT2002} together with a novel numerical
technique for simulating the ground state in the framework of the loop
cluster QMC method~\cite{Evertz2003, TodoK2001, Todo2004}, we show
that the difficulties mentioned above can be overcome.  Thus, we
successfully establish the quantitative ground-state phase diagram.

We start with the following Hamiltonian for the AF
Heisenberg chain
\begin{equation}
 \label{eqn:hamiltonian}
 {\cal H} = \sum_{j=1}^{L} J_j \, {\bf S}_j \cdot {\bf S}_{j+1} \,.
\end{equation}
Here, ${\bf S}_j$ is the spin-1 operator at site $j$ and $L$ the system
size; periodic boundary conditions are imposed.

For the bond-alternating model without disorder, where the coupling
constants $\{J_j\}$ are given by $J_j = 1 - (-1)^j \delta$
parameterized by the strength of bond alternation
$\delta$, its ground state has been
discussed in terms of the valence-bond solid (VBS)
picture~\cite{AffleckKLT1987}.  For the spin size $S$, the pattern of the
valence bonds $(m,n)$, where $m$ ($n=2S-m$) denotes the number of
effective singlet pairs on odd (even) bonds, changes from $(0,2S)$
to $(2S,0)$ successively as $\delta$ is increased from $-1$ to 1,
indicating the existence of $2S$ quantum phase
transitions~\cite{Oshikawa1992}.
Each VBS state has a topological hidden order, which is characterized
by the string order parameter~\cite{deNijsR1989}.

On the other hand, Affleck and Lieb studied Haldane's conjecture
on the basis of the Lieb-Schultz-Mattis (LSM) argument~\cite{AffleckL1986}.
Although the association between the VBS picture and the LSM argument has not been
fully understood for a long time, Nakamura and Todo have recently
shown that the ground-state expected value of the unitary operator
appearing in the LSM argument, given by
\begin{equation}
  z_L = \langle \exp [ {\rm i} \frac{2\pi}{L} \sum_{j=1}^{L} j S_j^z ] \rangle \,,
  \label{eqn:zl}
\end{equation}
functions as an order parameter, which characterizes the VBS
states~\cite{NakamuraT2002}.  The unitary operator in
eq.~(\ref{eqn:zl}) rotates the spins about the $z$-axis with the
relative rotation angle $2\pi/L$; thus, it generates a low-lying
excited state with an excitation energy of ${\cal O}(L^{-1})$.  Since the
twist order parameter~(\ref{eqn:zl}) measures the overlap between the
ground state and such a twisted excited state, $|z_L| \ne 1$ in the
thermodynamic limit evidences the existence of gapless low-lying
excitations or a degeneracy in the ground state.  Furthermore, it is
shown that in the $(m,n)$ VBS phase, $z_L$ converges to $(-1)^m$ for
$L \rightarrow \infty$.
We will see below that the twist order parameter works fairly well
even in the presence of disorder.

In what follows, we consider two different random distributions for
the couplings $\{J_j\}$ in eq.~(\ref{eqn:hamiltonian}).  The first one
is the {\em uniform distribution}, where the coupling constants are
distributed uniformly according to
\begin{equation}
\label{eqn:dist1}
P(J_j) = \begin{cases}
  1/2W & \text{if $|J_j - 1 + (-1)^j \delta| \le W$} \\
  0 & \text{otherwise.}
\end{cases}
\end{equation}
Here $0 \le W \le 1 - |\delta|$ must be fulfilled, otherwise
ferromagnetic bonds could appear in the system.  The second
distribution is given by
\begin{equation}
\label{eqn:j2}
J_j = [1 - (-1)^j \delta ] \, t_j \,
\end{equation}
with the quenched random numbers $t_j$ obeying the {\em power-law
distribution}~\cite{DasguptaM1980, Fisher1994, YangH2000}
\begin{equation}
\label{eqn:dist2}
P(t_j) = \begin{cases} R^{-1} t_j^{-1+1/R} & \text{if $0 < t_j \le 1$}
  \\ 0 & \text{otherwise,}
\end{cases}
\end{equation}
with a non-negative parameter $R$, where the $R\rightarrow0$ limit
corresponds to the nonrandom case ($t_j = 1$ for all $j$).  Note that
at $\delta=0$ the uniform distribution~(\ref{eqn:dist1}) with
$W=1$ is equivalent to the power-law distribution [eqs.~(\ref{eqn:j2}) and
  (\ref{eqn:dist2})] with $R=1$ besides
a trivial scaling factor; $J_j$'s are distributed uniformly between 0
and a finite cutoff.  In the following simulations, we take the
random average over 1000 samples for each parameter set.

The present model~(\ref{eqn:hamiltonian}) can be simulated efficiently
by the loop cluster QMC method~\cite{Evertz2003, TodoK2001} even in
the presence of randomness.  However, it should be pointed out that
the loop cluster method, which is based on the Suzuki-Trotter
path-integral representation, works indeed at a finite temperature.
Since the ground-state properties are only our main concern in the present
study, an effective extrapolation scheme, which we will explain below,
for taking the zero-temperature limit is essential.

We notice the fact that the ground state of the nearest-neighbor
AF Heisenberg chain of finite and even number of spins
is singlet, and there is a finite gap above the ground state.  In the
path-integral representation, the inverse of the gap is given by the
correlation length along the imaginary-time axis.  Since loop size
is directly related to the correlation length in real space as
well as in the imaginary-time direction~\cite{Evertz2003, TodoK2001}, the
system cannot distinguish whether the temperature is finite or zero,
if no loops wrap around the lattice in the imaginary-time direction.
In other words, the winding number of the loops in the imaginary-time
direction can be used as a good measure of the convergence to the
ground state.

Although there exist several means of implementing the above idea as a
ground-state QMC algorithm~\cite{Todo2004, EvertzL2001}, we employ the
following in the present study.  We start with a certain temperature.
During the thermalization Monte Carlo sweeps, the winding number of
the loops is monitored.  If one or more loops wrap around the system
in the imaginary-time direction, we double the inverse temperature.
This procedure will automatically adjust the simulation temperature so
that the system will be at the ground state effectively.

\begin{figure}[tb]
  \epsfig{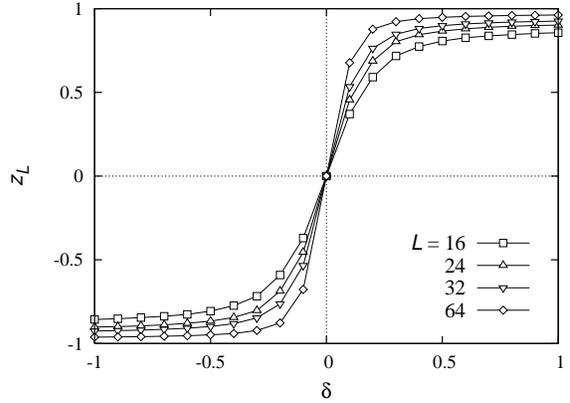}
  \caption{$\delta$-dependence of twist order parameter $z_L$ for
    spin-$\frac{1}{2}$ system with $R=0.5$.  At $\delta=0$, $z_L$ is
    zero irrespective of system size, while it converges to $\pm
    1$ for $\delta \ne 0$.} \label{fig:zl-spin-h}
\end{figure}

Before jumping to the spin-1 system, we discuss briefly the phase
diagram of the spin-$\frac{1}{2}$ system, for which the effects of
disorder on this system have been well established.  The ground state
of the non-bond-alternating spin-$\frac{1}{2}$ chain without disorder is critical.
By introducing infinitesimal randomness, the system is driven to
the RS phase, where there is also no excitation gap, but the
correlation function decays with an exponent different from that of the
nonrandom system~\cite{Fisher1994}.  The RS phase is characterized by
an infinite dynamical exponent, i.e., a logarithmic scaling of the
length and energy scales.  As a result, the uniform susceptibility
diverges as $\chi \sim 1 / T \log^2 T$ at low
temperatures~\cite{TodoKT1999}.

The RS phase is unstable against bond alternation.  The real-space
correlation becomes short-ranged immediately, though the spin gap
remains absent up to a finite strength of bond
alternation~\cite{HymanYBG1996, TodoKT1999}.  This phase is referred
to as the quantum Griffiths (QG) phase, where the uniform
susceptibility obeys the power law ($\chi\sim T^{-\gamma}$) at low
temperatures with a nonuniversal exponent $\gamma$ varying with
$\delta$.

In Fig.~\ref{fig:zl-spin-h}, the twist order parameter is plotted as a
function of $\delta$ for the spin-$\frac{1}{2}$ chain with $R=0.5$
(power-law distribution).  The twist order parameters with different
system sizes clearly cross at $\delta = 0$.  Note that in the random
system, the translational and parity symmetries are both broken in
each sample, and thus $z_L$ does not necessarily become zero at
$\delta = 0$.  However, one sees in Fig.~\ref{fig:zl-spin-h} that the
symmetries are restored after the random average is taken.  For a
nonzero $\delta$, the twist order parameter rapidly converges to $\pm
1$, though gapless QG phases extend on the both sides of the
RS point~\cite{HymanYBG1996, TodoKT1999}.  The present results
demonstrate clearly that the twist order parameter $z_L$ is not
affected by QG singularity, and thus it is an effective tool for
analyzing RS criticality.

\begin{figure}[tb]
  \epsfig{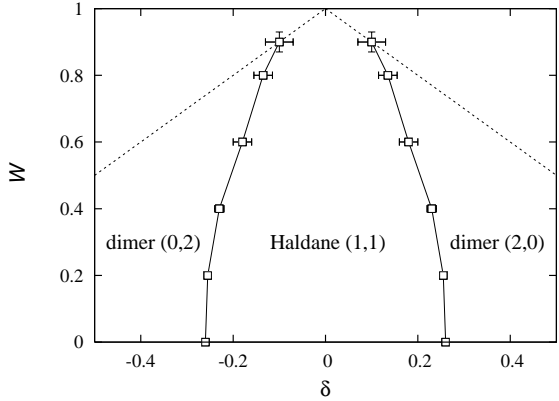}
  \caption{Ground-state phase diagram of spin-1 chain with
    uniform random-bond distribution.  Along the $\delta=0$ line, the
    Haldane phase survives up to $W=1$.} \label{fig:phase_box}
\end{figure}

In contrast to the spin-$\frac{1}{2}$ chain, the non-bond-alternating spin-1
Haldane system without disorder has a finite gap and a finite
correlation length~\cite{Haldane1983}.  In the previous DRG studies~\cite{HymanY1997, MonthusGJ1997},
it is predicted that the Haldane state is stable against a weak
disorder, while there occurs a quantum phase transition to the spin-1
RS phase at a critical strength of randomness.  In the previous QMC
analysis~\cite{TodoKT2000} of the model with a uniform random-bond
distribution~(\ref{eqn:dist1}), in which the uniform susceptibility
and the string order parameter were mainly investigated along the
$\delta=0$ line, it was concluded that a quantum phase transition
occurs at $W \simeq 0.95$ from the Haldane phase to the RS phase.  In
the present calculation, however, the twist order parameter decreases
with increasing system size in the entire range of $W$ ($0 \le W
\le 1$), and tends to converge to -1 without showing any crossing,
which indicates that the Haldane [(1,1) VBS] phase is stable in the
entire range of $W$.

This can be seen more clearly in the $\delta$-$W$ phase diagram shown
in Fig.~\ref{fig:phase_box}.  The phase boundaries are obtained from
the crossing point of the twist order parameter with different system
sizes ($L=8\cdots64$).  For small $\delta$'s, where the Haldane phase
existing at $|\delta| < 0.25997(3)$ for $W=0$~\cite{NakamuraT2002}
decreases gradually, the phase diagram [Fig.~\ref{fig:phase_box}] agrees
qualitatively with the one predicted by the DRG
analysis~\cite{Damle2002}.  However, the phase boundary
between the Haldane (1,1) and the dimer (2,0) phases (solid line)
merges with the parameter boundary $\delta + W = 1$ (dashed line) at
$\delta \simeq 0.1$, and does {\em not} reach $\delta = 0$ even at
$W=1$, indicating that there is no spin-1 RS phase in the model with the
uniform random-bond distribution.

Next, we examine the other random-bond distribution, i.e., the
power-law distribution [eqs.~(\ref{eqn:j2}) and (\ref{eqn:dist2})].
As already mentioned, the power-law distribution with $R=1$ is
equivalent to the uniform one with $W=1$; thus, it is expected that
the Haldane phase is stable at least up to $R=1$ also for the former
case.  However, for the power-law distribution, one can consider
a further strong disorder ($R>1$), i.e., a wider distribution on the
logarithmic scale, by which the Haldane phase might be broken~\cite{YangH2000}.

\begin{figure}[tb]
  \epsfig{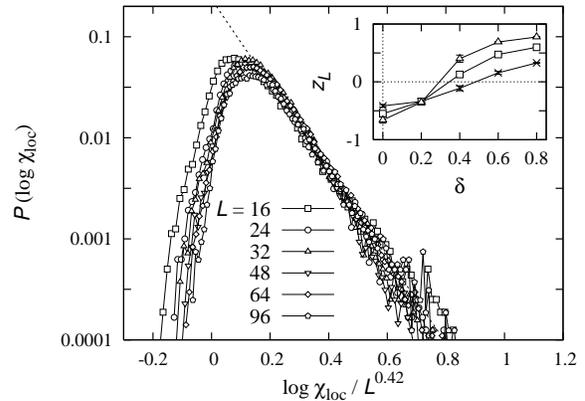}
  \caption{Scaling plot of distribution function of local
    susceptibility at RS point $(R,\delta)=(0.5,0.2)$.  Inset:
    $\delta$-dependence of twist order parameter for $R=0.5$
    (crosses, squares, and triangles for $L=8$, 16 and 32,
    respectively.)} \label{fig:xloc}
\end{figure}

In the inset of Fig.~\ref{fig:xloc}, the twist order parameter is
plotted as a function of $\delta$ in the weak randomness regime
($R=0.5$).  For $\delta > 0.2$, the twist
order parameter increases as the system size increases and tends to
converge to +1.  We identify this phase as the dimer (2,0) phase.  On
the other hand, $z_L$ tends to converge to -1 for $\delta < 0.2$,
indicating the Haldane (1,1) phase.

\begin{figure}[tb]
  \epsfig{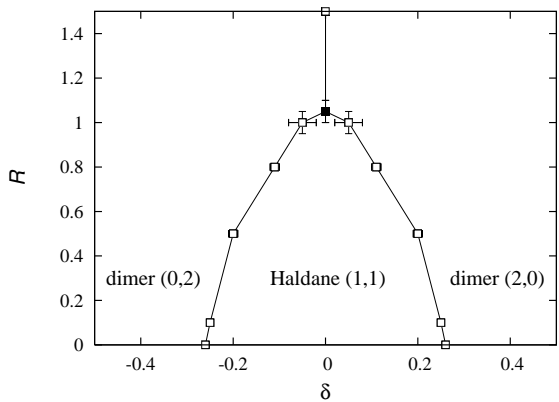}
  \caption{Ground-state phase diagram of spin-1 chain with
    power-law random-bond distribution.  The multicritical point is
    indicated by the filled square.}
  \label{fig:phase_power}
\end{figure}

At the crossing point $\delta = 0.20(1)$, a quantum phase transition
occurs, and the transition is expected to belong to the
spin-$\frac{1}{2}$ RS universality class~\cite{Damle2002}.
To confirm this prediction, we measured the distribution of
the local susceptibility
\begin{equation}
 \label{eqn:xloc}
  \chi_{{\rm loc},j} = \beta \langle m_j^2 \rangle
= \int_0^\beta \!\! d\tau \langle S_j^z(0) S_j^z(\tau) \rangle
\end{equation}
at the critical point $(R,\delta)=(0.5,0.2)$.  As seen in
Fig.~\ref{fig:xloc}, the distribution function of the logarithm of
the local susceptibility is scaled fairly well by assuming a
logarithmic scaling form, $P(\log \chi_{{\rm loc}}) \simeq \tilde{f}(\log \chi_{{\rm loc}} / L^\psi)$
with $\psi=0.42$.  This is consistent with the previous DRG prediction
for the RS phase~\cite{Fisher1994}, though the exponent
$\psi$ is slightly smaller than the predicted value ($\psi=1/2$).
This is additional support to the validity of applying the twist order parameter to 
randomness-driven quantum phase transitions.  Repeating similar
analyses, we obtain the entire $\delta$-$R$ phase diagram of the random
Haldane chain with the power-law
distribution~[Fig.~\ref{fig:phase_power}].

\begin{figure}[tb]
  \epsfig{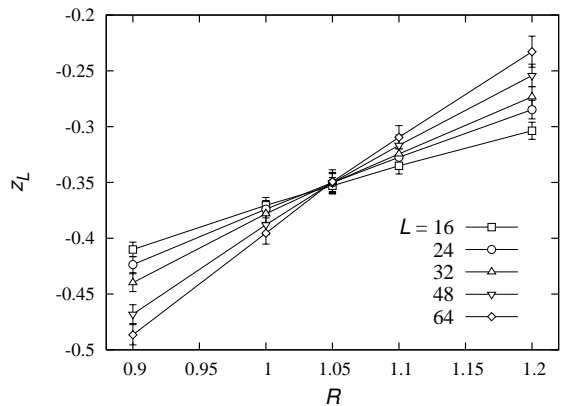}
  \caption{$R$-dependence of twist order parameter $z_L$ of
    spin-1 system with $\delta=0$.  The data with different system
    sizes cross at $R\simeq 1.05$.} \label{fig:zl_tricritical}
\end{figure}

Although the phase diagram for the power-law distribution is similar
to Fig.~\ref{fig:phase_box} for small $R$'s, the overall shape
of the phase boundaries indicates the existence of a multicritical
point, where two critical lines merge with each other at a finite
$R$.  To locate the multicritical point, we calculate $z_L$
for several system sizes ($L=16\cdots64$) along the $\delta=0$ line.
The results for $0.9 \le R \le 1.2$ is shown in
Fig.~\ref{fig:zl_tricritical}, where the data with different system
sizes clearly cross at $R_{\rm c} \simeq 1.05$.  Thus, we conclude
that there exists a multicritical point at $(R,\delta)=(1.05,0)$, which is
indicated by the solid square in Fig.~\ref{fig:phase_power}.  Below the
multicritical point, the Haldane phase survives, though the spin gap
vanishes at a certain $R$ ($<R_{\rm c}$), where a crossover from the
gapped Haldane phase to the gapless Haldane (or QG) phase occurs.  In
the case of a uniform distribution~(\ref{eqn:dist1}), the crossover is observed at $W \simeq
0.7$~\cite{TodoKT1999}, though we have not yet examined it for the
power-law distribution.  For $R > R_{\rm c}$, on the other hand, the
twist order parameter is expected to converge to a nontrivial value (i.e.,
$z_L \neq \pm 1$) in the thermodynamic limit, where the spin-1 RS phase is
realized~\cite{HymanY1997, MonthusGJ1997,
  Damle2002}.

To summarize, we reported the results of our QMC simulations on the
bond-alternating random Haldane chain.  By introducing the
ground-state loop cluster QMC method and the twist order parameter, we
have successfully calculated the ground-state phase diagram.
In particular, we demonstrated that the twist order parameter, introduced
originally for pure spin chains, is also effective for random
spin chains.  Indeed, it is shown that the behavior of the twist order
parameter observed in the present study can be discussed more directly
in terms of the numerical DRG approach, in which one can calculate the
topological order parameter for an approximate VBS-like ground state
explicitly~\cite{TodoAT2004}.

For the uniform distribution, the present result, i.e., the absence of
the spin-1 RS phase, does not agree with the previous
finite-temperature QMC result, in which a multicritical point was
suggested~\cite{TodoKT2000}.  The possible reason for this disagreement is
that the finite-temperature QMC method might easily fail to take into
account rare and low-energy-scale but very strong correlations, which
are essential in random spin systems.  In contrast, in the present
ground-state algorithm, simulation temperature is automatically
adjusted according to the magnitude of the gap of each random sample,
so that the physical quantity at the zero temperature is calculated at
an optimal cost.  This algorithm is useful in
simulating not only random systems but also those without disorder~\cite{Todo2004}.

For the power-law distribution, on the other hand, we established a
phase diagram with a multicritical point, whose location was also
determined accurately using the twist order parameter.  The present
phase diagram agrees qualitatively with the recent DRG
prediction~\cite{Damle2002}, though the numerical
confirmation of spin-1 RS criticality, which is expected to
realize in the strong disorder regime ($R>R_{\rm c}$), still remains
as a future problem.

We acknowledge our fruitful discussion with M. Nakamura.  The simulations
presented in the this Letter have been performed using the facility of
the Supercomputer Center, Institute for Solid State Physics,
University of Tokyo.  S.T. acknowledges the support by a Grant-in-Aid for
the Scientific Research Program (No.15740232) from the Japan Society for
the Promotion of Science.

\end{document}